\documentclass[prb,twocolumn,showpacs]{revtex4}  
\usepackage{graphicx}  
  
\newcommand{\be}{\begin{equation}}  
\newcommand{\ee}{  \end{equation}}  
\newcommand{\ba}{\begin{eqnarray}}
\newcommand{\ea}{  \end{eqnarray}}  
  
\begin{document}  
  
\title{Statistical fluctuations of pumping and rectification currents 
in quantum dots} 
  
\author{M. Mart\'{\i}nez-Mares and C. H. Lewenkopf}  
   
\affiliation{Instituto de F\'{\i}sica, Universidade do Estado do  
               Rio de Janeiro, R. S\~ao Francisco Xavier 524,  
               20550-900 Rio de Janeiro, Brazil}  
  
\author{E. R. Mucciolo}  
  
\affiliation{Department of Physics, Duke University, Durham, North  
               Carolina 27708 \\ Departamento de F\a'{\i}sica,  
               Pontif\a'{\i}cia Universidade Cat\'olica do Rio de  
               Janeiro, C.P. 38071, 22452-970 Rio de Janeiro, Brazil}  
  
\date{\today}  
   
\begin{abstract}  
We investigate the statistical fluctuations of currents in chaotic
quantum dots induced by pumping and rectification at finite
temperature and in the presence of dephasing. In open quantum dots, dc
currents can be generated by the action of two equal-frequency ac gate
voltages. The adiabatic regime occurs when the driving frequency is
smaller than the electron inverse dwell time. Using numerical
simulations complemented by semiclassical calculations, we consider
both limits of small and large number of propagating channels in the
leads when time-reversal symmetry is fully broken. We find that at
intermediate temperature regimes, namely, $k_BT \alt \Delta$, where
$\Delta$ is the mean single-particle level spacing, thermal smearing
suppresses the current amplitude more effectively than
dephasing. Motivated by recent theoretical and experimental works, we
also study the statistics of rectified currents in the presence of a
parallel, Zeeman splitting, magnetic field.
\end{abstract}   
  
   
   
\maketitle

\section{Introduction}
\label{sec:intro}

The study of non-equilibrium transport phenomena in mesoscopic 
electronic systems has attracted much attention in recent years, in 
particular for ballistic quantum dots formed in semiconductor 
heterostructures. Besides the investigation of non-linear $I-V$ 
characteristics, it has been proposed that dc currents can be induced 
by the simultaneous application of two ac, shape-deforming gate 
voltages, the so-called pumping 
effect.\cite{Thouless83,Spivak95,Brouwer98,Zhou99} An experimental 
realization of an adiabatic quantum dot pump of electrons under these 
conditions has been reported.\cite{Switkes99} For open quantum dots, 
where the system is connected to reservoirs by one or more propagating 
channels at each contact, the pumping current has usually both pure 
quantum (dissipationless) and rectified (dissipative) 
components.\cite{Brouwer01} While attempts have been made to maximize 
the former, it seems that current experiments are still dominated by 
the latter. Nevertheless, the effects of phase coherence can also be 
present in the rectified component. A clear indication of that has 
been given by the observation of a mesoscopic spin current 
\cite{Mucciolo02} by Watson {\it et al.} \cite{Watson03} in a setup 
where most likely rectification is still the primary source of 
pumping. 
  
In electronic mesoscopic systems, phase coherence and the consequent 
lack of self-averaging are responsible for the sample-to-sample 
fluctuations of most transport properties. For systems with an 
underlying chaotic dynamics, these fluctuations are characterized by 
universal statistics. Therefore, for either quantum pumping or 
rectification, it is important to understand what are the universal 
features of current fluctuations. Being universal, these features can 
be described, in principle, by a theory based only on a few hypotheses 
about the system symmetries and its connection to the reservoirs and 
the environment. In fact, by comparing universal probability 
distributions and correlation functions with actual experimental 
measurements, one may be able to gain information about nonuniversal 
effects. Also, through this procedure, it might be possible to 
estimate the amplitude of phase decoherence present in the device. 
  
The theory of zero-temperature statistical properties of the pumping 
current in quantum dots \cite{Brouwer98} explains qualitatively 
several aspects of the experiment.\cite{Switkes99} A quantitative 
approach, however, requires treating both temperature and the 
consequent dephasing effects. Previous work addressing dephasing did 
not account for the thermal rounding of the Fermi 
surface.\cite{Moskalets01,Cremers02} Albeit the thermal convolution 
for the pumping current being casted in a very simple expression, 
assessing the fluctuations requires the computation of nontrivial 
energy correlation functions. These correlation functions were 
calculated only in the absence of dephasing and when a large number of 
propagating channels ($N \gg 1$) is 
present.\cite{Shutenko00,Vavilov01b} In this limiting case, the role 
of temperature is well understood. In the preset work we use 
analytical and numerical calculations to study the effects of 
temperature and dephasing in the case of small number of channels $N 
\ge 2 $, which is more relevant to the experiments. Also, we revisit 
the case of $N\gg 1$ from a semiclassical viewpoint, where results can 
be obtained in a more transparent and insightful way than from the 
diagrammatic technique. 
 
The paper is organized as follows. In Sec. \ref{sec:theory}, we 
briefly present the general theory of pumping currents in open QDs in 
the adiabatic regime. Finite-temperature pumping current fluctuations 
are expressed as integrals over correlation functions of scattering 
matrices in Sec. \ref{sec:thermal}. In Sec. \ref{sec:semiclasSmatcor}, 
we show that such correlation functions are readily calculated by 
means of the semiclassical approximation. The small-$N$ limit is 
investigated numerically in Sec. \ref{sec:smallN}. The combined effect 
of dephasing and temperature is discussed in 
Sec. \ref{sec:dephasing-standard-pumping}. In 
Sec. \ref{sec:rectification} we address the influence of capacitive 
couplings on the pumping currents and their rectification 
effects. Mesoscopic fluctuations of spin pumping currents are 
discussed in Sec. \ref{sec:spin}. Finally, in 
Sec. \ref{sec:conclusions}, we present our conclusions.

\section{Pumping currents}   
\label{sec:theory}   
We consider a quantum dot (QD) formed by laterally confining a region 
of a two-dimensional electron gas (2DEG) through multiple 
electrostatic gate voltages. The electrons in the QD have free access 
to the rest of the 2DEG through two point-contact leads. We assume 
that the ``right" and ``left" leads support $N_R$ and $N_L$ fully 
transmitting modes, respectively. The confining potential of the QD 
undergoes a periodic shape deformation caused by the ac modulation of 
two gate voltages, which can be parameterized by $X_1(t) = 
a_1\cos(\omega t + \phi_1)$ and $X_2(t) = a_2 \cos(\omega t + 
\phi_2)$. Here, $\omega$ denotes the driving frequency, $\phi= 
\phi_2-\phi_1$ is the (constant) phase difference between voltages, 
and $a_1$ and $a_2$ are their amplitudes. We focus our treatment of 
the problem to the adiabatic regime, when the driving frequency of the 
perturbations is much smaller than the electron escape rate (or 
inverse dwell time): $\omega \ll \Gamma/\hbar = N\Delta /2\pi\hbar$, 
where $N = N_R + N_L$ and $\Delta$ is the mean level spacing in the QD 
when isolated from the reservoirs. 
 
The adiabatic deformations of the QD confining potential induce charge 
transfer through the point contacts. In the linear response limit, the 
amount of charge passing through a given contact $l$ ($l=L,R$) is 
expressed as\cite{Buttiker94} 
\be  
\label{eq:dQ} 
\delta Q_l = e \left( \frac{dn_l}{dX_1} \delta X_1 + 
\frac{dn_l}{dX_2} \delta X_2 \right),  
\ee 
where the so-called emissivities can be written in terms of scattering 
matrix elements, namely, 
\be 
\label{eq:emissivity} 
\frac{dn_l}{dX_i} = \frac{1}{2\pi} \sum_\beta \sum_{\alpha \in l} 
\text{Im} \left( \frac{\partial S_{\alpha\beta}} {\partial X_i} 
S^\ast_{\alpha\beta} \right), 
\ee 
for $i=1,2$. Thus, the total charge per cycle transfered through the 
contact $l$ can be evaluated by integrating Eq. (\ref{eq:dQ}) over one 
period, 
\be 
\label{eq:current} 
Q_l = e \int_0^{2\pi/\omega} dt\, \left[ \frac{dn_l}{dX_1} 
\frac{dX_1}{dt} + \frac{dn_l}{dX_2} \frac{dX_2}{dt} \right]. 
\ee 
One can use Eqs. (\ref{eq:emissivity}), (\ref{eq:current}), and 
rewrite the integral over time as an integral over the surface area 
${\cal A}$ swept by $X_1$ and $X_2$ in parameter space over one 
period. Moreover, it is not difficult to prove that $Q_L = - Q_R$, 
since charge is not accumulated inside the QD (see below). As a 
result, we can cast the zero temperature pumping current going from 
left to right reservoirs as \cite{Brouwer98} 
\be   
\label{eq:defI0}   
I_0(E_F) = \frac{\omega Q_L}{2\pi} = \frac{e\omega}{2\pi} 
\int_{\cal{A}} \!dX_1 dX_2 \,\, \Pi_0^L (E_F, {\bf X}), 
\ee 
where ${\bf X} = (X_1,X_2)$. Notice that only electrons in states at 
the Fermi energy contribute to the current (we have used the subscript 
in the current to indicate temperature). In terms of the $S$ matrix, 
the pumping response function $\Pi_0^l$ reads 
\be  
\label{eq:defPi0} 
\Pi_0^l(E,{\bf X}) = \frac{1}{\pi}\sum_\beta\sum_{\alpha \in l} 
\,\mbox{Im}\!\left(\frac{\partial S_{\alpha\beta}^*}{\partial X_1} 
\frac{\partial S_{\alpha\beta} }{\partial X_2}\right). 
\ee 
The energy and parameter dependences of the response function are 
determined by those of the scattering matrix. Therefore, it is 
necessary to make an explicit connection between the latter and the 
microscopic details of the system. Such connection is provided by 
\cite{Mahaux69} 
\be  
\label{eq:S}   
S(E,{\bf X}) = \openone - 2i\pi W^\dagger \frac{1}{E - H({\bf X}) + i 
\pi W W^\dagger}W, \ee 
where $E$ is the energy of incoming and outgoing electrons (assuming 
that no bias voltage is applied to the system). $W$ is a $M\times N$ 
matrix that gives the coupling between the $M$ resonant 
single-particle electronic states in the QD, described by the $M\times 
M$ Hamiltonian matrix $H({\bf X})$, and the $N$ propagating modes in 
the leads. 
 
Throughout this paper we work with the hypothesis that the coupling 
between the QD and the leads is maximal, meaning that there are no 
barriers for electrons entering and leaving the system. In this 
picture, $W$ becomes independent of any external parameter. This 
hypothesis is good inasmuch the changes in the confining potential due 
the periodic perturbation are local and take place far for the QD 
openings. This means that the distance $\ell$ between the openings and 
the perturbed region must be such that $\ell/\lambda_F\gg 1$. Should 
that not be the case, then $dW/dX\neq 0$, resulting in an additional 
pumping mechanism similar to that of a classical peristaltic pump, 
where transport occurs because the constriction are periodically 
opened and closed while the internal potential is varied. Although the 
existence of such classical effect in the experiments cannot be 
entirely ruled out, we do not take it into account in our 
analysis. The clear existence of mesoscopic, sample-to-sample 
fluctuations and a marked temperature dependence of the magnitude of 
the current amplitude in the experiments seems to indicate that the 
peristaltic effect can be made rather weak. Thus, for a 
parametrically constant $W$, we have 
\be  
\label{eq:dSdX}   
\frac{\partial S}{\partial X_i} = - 2\pi i W^\dagger D^{-1} 
\frac{\partial H}{\partial X_i} D^{-1} W 
\ee 
where $D = E - H({\bf X}) + i \pi W W^\dagger$. Using 
Eq. (\ref{eq:dSdX}), it is straightforward to show that $\Pi\equiv 
\Pi^L=-\Pi^R$. 
   
We are interested in chaotic quantum dots, where the Hamiltonian $H$
can be modeled as a member of one of the Gaussian ensembles of random
matrices.\cite{Mehta91} For this case, the Hamiltonian matrix elements
are assumed uncorrelated but equally distributed. Their variance
$\lambda^2/M$ determines the mean resonance spacing at the center of
the band, $\Delta=\pi\lambda/M$. Several studies support that, in
general, the elements of $\partial H/ \partial X$ have themselves also
Gaussian random entries (see, for instance, Ref.~\onlinecite{Barth99}
and references therein). We choose their variance to be
$(\lambda/MX_c)^2$.\cite{comment} In this way, we set the scale of
$X$, making $X=X_c$ correspond to the average parametric change
necessary to cause one level crossing. In other words, $X_1$ and $X_2$
are measured in units of the average parametric level
crossing. Motivated by the fact that pumping currents are usually
generated in the presence of an external perpendicular magnetic field
that breaks time-reversal symmetry, in what follows we only address
the case where $H$ belongs to the unitary ensemble (GUE).
 
The statistical theory presented above is an alternative to the 
maximal entropy approach used to calculate the pumping current 
fluctuations in Refs.~\onlinecite{Brouwer98,Cremers02}. At $T=0$ both 
approaches are equivalent. However, while the maximum entropy has the 
advantage of leading to analytical expressions for the distribution of 
$\Pi_0$ at $T=0$, it cannot be consistently extended to finite 
temperatures. \cite{Alves02} As we show bellow, temperature plays a 
very important role in suppressing the pumping current fluctuations.

\subsection{Thermal fluctuations} 
\label{sec:thermal}  
 
Let us start addressing the thermal rounding of the Fermi surface,
postponing the discussion of dephasing to
Sec. \ref{sec:dephasing-standard-pumping}. The thermal smearing is
easily accounted for by the integral
\be  
\label{eq:varI} 
I_T(\mu) = \int dE \left(- \frac{\partial f_T} {\partial E} 
\right) I_0(E), 
\ee 
where $f_T = \{\exp[(E - \mu)/T] + 1\}^{-1}$ (we assume $k_B=1$ 
hereafter). Hence, the pumping current variance reads 
\be  
\label{eq:varthermal} 
\langle I_T^2 \rangle = \left( \frac{e\omega T}{2\pi} \right)^2 
\!\int_{-\infty}^{\infty} \!\!d \varepsilon \frac{d}{dT}\!\left[ 2 T 
\,\mbox{sinh}\! \left( \frac{\varepsilon}{2T} \right) \right]^{-2} 
\!C_0(\varepsilon),  
\ee 
where $C_0$ is the pumping response autocorrelation function at $T=0$,
defined by
\be 
\label{eq:defC0}  
C_0(\varepsilon) = \int_{\cal A} dX_1\, dX_2 \int_{\cal A} dY_1\, dY_2 
\ D(\varepsilon, {\bf X} - {\bf Y})  
\ee 
and 
\be 
\label{eq:defD}  
D(\varepsilon,  {\bf X} - {\bf Y}) =  
\left\langle \Pi_0 \left(\mu +\frac{\varepsilon}{2}, {\bf X}\right)  
\Pi_0 \left(\mu -\frac{\varepsilon}{2}, {\bf Y} \right)\right\rangle. 
\ee 
Here $\langle \cdots\rangle$ indicates ensemble averaging. If the
amplitude of the periodic perturbations is sufficiently weak, such
that $\sqrt{\cal A} \ll X_c$, we can approximate the surface integrals
by the mean value of the response function at $X=0$ times the area,
yielding
\be  
\label{eq:defC} 
C_0(\varepsilon) \approx {\cal A}^2 D(\varepsilon, 0). 
\ee 
At very low temperatures, the variance of the pumping current becomes
proportional to $C_0(0)$, i.e., $\langle \Pi_0^2 \rangle$. In the
absence of dephasing, we find that $\Pi_0$ has zero mean and variance
given by (see Appendix \ref{sec:appDeriva_var(Pi0)})
\be \label{eq:varPi0} \text{var} \, \Pi_0 = \frac{16}{\pi^2} \frac{N_L \, 
N_R} {\left( N^2-1 \right) \left( N^2-4 \right)}. 
\ee 
Note that var($\Pi_0$) diverges for $N=2$. As $N$ increases, this
divergence is smeared out. For large $N$, $C_0(\varepsilon)$ quickly
tends to a universal, ``semiclassical" form. The derivation of this
function is presented below.

\subsection{Semiclassical pumping response autocorrelation function} 
\label{sec:semiclasSmatcor}  
 
When the number of open channels $N\gg 1$, the pumping response
autocorrelation function of Eq. (\ref{eq:defC0}) can be directly
evaluated using Miller's semiclassical $S$-matrix,\cite{Miller75}
namely,
\be  
\label{eq:Miller}  
\widetilde{S}_{\alpha\beta}(E,{\bf X}) = \sum_{\mu(\alpha \leftarrow 
\beta)} \sqrt{p_\mu(E,{\bf X})} \, e^{i\sigma_\mu(E,{\bf X})/\hbar} \;,  
\ee 
where the classical trajectories that start at channel $\beta$ and end
at channel $\alpha$ are labeled by $\mu(\alpha \leftarrow \beta)$.
Accordingly, $\sigma_\mu$ is the reduced action (with a Maslov phase
included) and $p_\mu$ is the classical transition probability for
going from $\beta$ to $\alpha$ following the path $\mu$. (Quantities
indicated by a wide tilde are obtained in the semiclassical
approximation.) In the derivation of Eq. (\ref{eq:Miller}), the
absence of tunneling barriers between the scattering and the
asymptotic regions is implicit.
  
Let us write the pumping response function as $\Pi_0 =
\pi^{-1}\sum_\beta \sum_{\alpha \in l} J_{\alpha\beta}$. The
semiclassical approximation for $J_{\alpha \beta}$ reads
\be  
\widetilde{J}_{\alpha\beta} (E,{\bf X}) = \sum_{\mu,\nu} \frac{\partial 
\sigma_\mu}{\partial X_1} \frac{\partial \sigma_\nu}{\partial X_2} 
\sqrt{p_\mu p_\nu} \, \sin\!\left(\!\frac{\sigma_\mu - \sigma_\nu}
{\hbar}\!\right), 
\ee 
where $\mu$ and $\nu$ are trajectories connecting the same pair of two
arbitrary channels $\alpha$ and $\beta$. The parametric action
derivatives are defined as \cite{Almeida98}
\be  
\label{eq:dsigmadX}  
Q_{i\mu} \equiv  
\frac{\partial\sigma_\mu}{\partial X_i}=  \int_0^{t_\mu} \! dt \,  
      \frac{\partial H}{\partial X_i}  
        \Big({\bf p}(t), {\bf q}(t), {\bf X}\Big) \;,  
\ee  
where the integral in evaluated along the trajectory $\mu$ over the
time $t_\mu$ it spends in the QD.
  
Our ergodic hypothesis is that averages over random matrix ensembles
are equivalent to the energy averages taken here. We average
$\widetilde{J}_{\alpha\beta} (E,{\bf X})$ over an energy window
$\delta E$ where the classical dynamics presents little changes,
nonetheless fulfilling $\delta E \gg \Delta$. For this task, as
customary, we neglect the energy dependence of the probabilities
$p_\mu$ and keep only contributions from diagonal terms. This
approximation is justified for trajectories with dwell times shorter
than the Heisenberg time $\tau_{\text{H}}\equiv h/\Delta$, since they
are, in general, uncorrelated for chaotic systems. Fortunately,
without barriers, trajectories with $t$ exceeding $\tau_H$ are
statistically negligible in the semiclassical regime of $N\gg 1$. In
the absence of system specific symmetries, the diagonal approximation
reads $\left\langle \exp[i(\sigma_\mu - \sigma_\nu)/\hbar]
\right\rangle_{\delta E} = \delta_{\mu \nu}$. It implies that $\langle
\widetilde{J}_{\alpha\beta} (E,{\bf X}) \rangle_{\delta E}=0$ and
hence $\langle \widetilde{\Pi}_0 (E,{\bf X}) \rangle_{\delta E}=0$.

We use a similar procedure to obtain the semiclassical pumping
response autocorrelation function $\widetilde{D}(\varepsilon,
\delta{\bf X})$, defined as in Eq.\ (\ref{eq:defD}). Now we deal with
a product of two $\widetilde{\Pi}_0$ functions evaluated at different
energies, $E\pm \varepsilon/2$, and parameter values, $\overline{\bf
X} \pm \delta {\bf X}/2$. The actions $\sigma_\mu (E \pm
\varepsilon/2, \overline{\bf X} \pm \delta {\bf X}/2)$ are
approximated in leading order of classical perturbation theory as
\be 
\sigma_\mu (E \pm \frac{\varepsilon}{2}, \overline{\bf X} \pm
\frac{\delta {\bf X}}{2}) = \sigma_\mu(E, \overline{\bf X}) \pm
t_\mu\frac{\varepsilon}{2} \pm {\bf Q}_\mu \cdot \frac{\delta {\bf
X}}{2}.
\ee

Let us start examining the $J$ autocorrelation function. The diagonal
approximation is used to compute the energy average of $\sin[
(\sigma_\mu - \sigma_\nu) /\hbar] \sin[ (\sigma_{\mu^\prime} -
\sigma_{\nu^\prime}) /\hbar]$ and gives
\ba  
\label{eq:Dexplicit}  
\left\langle \widetilde{J}_{\alpha\beta}^{(+)}
 \widetilde{J}_{\alpha^\prime\beta^\prime}^{(-)}
 \right\rangle_{\delta\varepsilon}=&&\!\!\!\!  \frac{1}{2\hbar^4}
 \text{Re} \!\!\!\!\sum_{{\mu,\nu(\alpha \leftarrow \beta)}
 \atop{\mu^\prime,\nu^\prime}(\alpha^\prime \leftarrow \beta^\prime)}
 (\delta_{\nu \nu^\prime}\delta_{\mu \mu^\prime}+ \delta_{\nu
 \mu^\prime}\delta_{\nu \mu^\prime}) \nonumber\\ &&
 \!\!\!\!\!\!\!\!\!\!\!\!\!\!\!\!\!\!\!\!\!\!\!\!\!\!\!\!\!\!\!\!\!\!
 \times\sqrt{p_\mu p_{\mu^\prime} p_\nu p_{\nu^\prime}}
 Q_{1\mu}Q_{1\mu^\prime}Q_{2\nu} Q_{2\nu^\prime} \\ \nonumber &&
 \!\!\!\!\!\!\!\!\!\!\!\!\!\!\!\!\!\!\!\!\!\!\!\!\!\!\!\!\!\!\!\!\!\!
 \times\exp\!\!\left\{-\frac{i}{\hbar}\Big[\varepsilon(t_\nu -t_\mu)+
 \delta{\bf X} \cdot({\bf Q}_\nu - {\bf Q}_\mu)\Big] \right\},
\ea
where $J^{(\pm)}_{\alpha\beta} \equiv
J_{\alpha\beta}(E\pm\varepsilon/2, \overline{\bf X} \pm {\bf X}/2)$.
Since no special attention is payed to time-reversal symmetric paths,
Eq.\ (\ref{eq:Dexplicit}) represents the semiclassical correlation
function for broken time-reversal symmetry.

We proceed using classical sum rules to convert the sums in Eq.\
(\ref{eq:Dexplicit}) into time integrals. For that purpose, orbits are
grouped with respect to common traversal times and averages are taken
within these sets. In what follows, we describe the details how this
procedure is implemented following two basic steps.
  
For chaotic systems the transition probabilities $p_\mu$ follow the
analogue of the Hannay-Ozorio de Almeida sum rule for open systems,
\cite{scattering-sum-rule}
\be  
\label{eq:sumrule}  
\sum_{t \le t_\mu \le t + \delta t} p_\mu = \frac{1}{N\tau}\,
\mbox{e}^{-t/\tau}\, \delta t \equiv \overline{p}(t)\delta t, 
\ee
where $\sum_{t \le t_\mu \le t + \delta t}\,p_\mu$ is the sum of all
classical transition probabilities following the trajectories $\mu$
belonging to a time interval $[t, t+\delta t]$, where $\delta t$ is
classically small. In Eq. (\ref{eq:sumrule}), the decay time is $\tau
= \hbar/\Gamma$,\cite{Vallejos01}, where $\Gamma$ is the escape width,
also known in this context as the Weisskopf $S$-matrix autocorrelation
length.
  
The transition probabilities $p_\mu$ and the parametric action
derivatives ${\bf Q}_\mu$ are uncorrelated. In addition, $Q_{1\mu}$
and $Q_{2\mu}$ are uncorrelated, provided $\mu$ dwells inside the QD
for a couple traversal times. Thus, for fixed $\overline{X}$, the
time-average in Eq.\ (\ref{eq:sumrule}) runs over a large number of
scattering orbits with ${\bf Q}_\mu(t)$, which may be considered as
samples of the probability distribution $P_t({\bf Q})=
P_t(Q_1)P_t(Q_2)$.  The latter is assumed to be Gaussian,
\cite{Almeida98}
\begin{equation}
P_t(Q_i) = \frac{1}{\sqrt{2\pi \overline{Q^2_i}(t)}} \exp \left[-
\frac{Q_i^2}{2\overline{Q^2_i}(t)} \right].
\label{eq:gaussantz}
\end{equation}
The Gaussian width is a function of $\overline{X}$ and $\overline{E}$,
known to grow diffusively with time,\cite{Goldberg91} namely,
\begin{equation}
 \overline{Q^2_i}(t) = \frac{1}{{\cal N}(t)} \sum_{t \le t_\mu \le t +
\delta t} Q^2_{i\mu} = B t ,
\end{equation}
where ${\cal N}(t)$ is the number of trajectories $\mu$ within the
time window $[t, t + \delta t]$. The diffusion constant $B$ is not
expected to depend on $i$ since both $X_1$ and $X_2$ are generic shape
parameter. (This is not the case if, for instance, $X_1$ would stand
for a magnetic fiend and $X_2$ for a shape deformation.) $B$ is given
by
\be  
\label{eq:B}  
B(E, {\bf X}) = 2\!\int_0^\infty \!\!\!dt \,\overline{ \frac{\partial 
H}{\partial X} \Big( {\bf p}(t),{\bf q}(t) \Big) \frac{\partial 
H}{\partial X} \Big( {\bf p}(0),{\bf q}(0) \Big) }. 
\end{equation}  
It is important to notice that the decay of the classical correlation
function in Eq. (\ref{eq:B}) need only be integrable (many available
chaotic systems do not exhibit full exponential decay of the
correlations). A detailed discussion of the classical properties of
${\bf Q}_\mu$ and its semiclassical implications for density
correlation function can be found in Ref. \onlinecite{Almeida98}. In
particular, there it is shown that $X_c$ is intimately related to the
classical diffusion constant, namely, $X_c =
(h\Delta/B)^{1/2}/\pi$.\cite{comment}
  
By taking the Gaussian average over ${\bf Q}_{i\mu}$, inserting
(\ref{eq:sumrule}) and (\ref{eq:B}) into (\ref{eq:Dexplicit}), and
integrating over time, we arrive at
\be \label{eq:sccorrfunction} \widetilde{D}(\varepsilon,{\bf \delta
X}) = \frac{{\rm var}\,
\widetilde{\Pi}_0}{\left\{\left(\frac{\varepsilon}{\Gamma}\right)^2 +
\left[1+ \frac{2(\delta {\bf X})^2}{N X_c^2}\right]^2\right\}^2}, \ee
with 
\be 
\label{eq:scvarPi} 
{\rm var}\,\widetilde{\Pi}_0 = \frac{4}{\pi^2 N^2} \frac{1}{X_c^4}.
\ee
This expression agrees with the random matrix theory results of
Ref. \onlinecite{Vavilov01b}, also obtained for $N\gg 1$. The
semiclassical ${\rm var}\,\widetilde{\Pi_0}$ is also consistent with
Eq.\ (\ref{eq:varPi0}), as it should.

\begin{figure}   
\includegraphics[width=9.0cm]{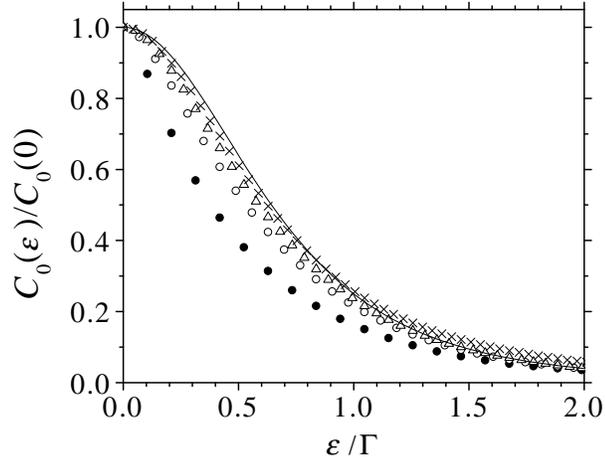}
\caption{Pumping current autocorrelation function $C_0(\varepsilon)/C_0(0)$ 
versus energy $\varepsilon$ in units of $\Gamma$. The solid circles represent 
the $N=N_L+N_R=4$ case, whereas empty circles, triangle, and crosses stand 
for $N=6$, $8$, and $10$, respectively. In all cases $N_L=N_R$. The solid 
line is the autocorrelation function given by Eq.\ (\ref{eq:sccorrfunction}).} 
\label{fig:ucor_scaled}   
\end{figure}   

\subsection{Small-$N$ limit} 
\label{sec:smallN}  
 
\begin{figure}   
\includegraphics[width=9.0cm]{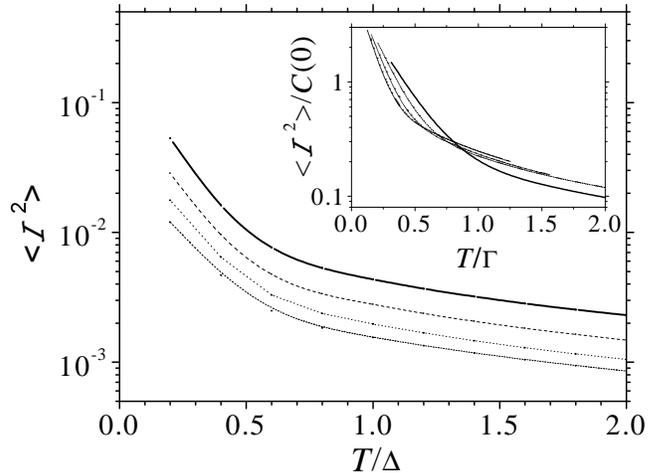}   
\caption{Pumping current variance $\langle {\cal I}^2 \rangle$ in 
units of $(e\omega {\cal A}/2\pi)^2$ versus temperature $T$ in units 
of $\Delta$. The thick solid line stands for the $N=N_L+N_R=4$ case, 
whereas dashed, dotted, and short-dashed lines stand for $N=6$, $8$, 
and $10$, respectively. In all cases $N_L=N_R$. Inset: Same cases, but 
for the pumping current variance scaled by $C_0(0)$ (temperature 
measured in units of $\Gamma$).} 
\label{fig:uvarpumpNT}   
\end{figure}

It is a difficult technical task to calculate an analytical expression 
for $C_0(\varepsilon)$ in the small-$N$ limit, which is the case of 
experimental interest. To bridge this gap, we relied on numerical 
simulations. We typically generated $10^5$ members of an ensemble of 
$S$ matrices defined by Eq. (\ref{eq:S}), following the prescription 
given in Ref. \onlinecite{Alves02}. The results for $N>2$ are 
illustrated by Fig. \ref{fig:ucor_scaled}. Notice that as $N$ 
increases, the semiclassical limit is attained very 
fast. Nevertheless, for $N=4$, which is the experimental case in Ref.\ 
\onlinecite{Switkes99}, the depart from the semiclassical limit is 
still significant. The effect of having a small $N$ is best captured 
in Fig.\ \ref{fig:uvarpumpNT}, where we show the pumping current 
variance in units of $(e\omega {\cal A}/2\pi)^2$, namely, $\langle 
{\cal I}^2 \rangle \equiv \langle {I_T}^2 \rangle/(e\omega {\cal 
A}/2\pi)^2$, as a function of temperature. The numerical results were 
obtained by carrying out the integration in Eq.\ (\ref{eq:varthermal}) 
through an interpolating of the data shown in Fig.\ 
\ref{fig:ucor_scaled}. We can observe that the fluctuations decrease 
with increasing temperature and $N$. In the inset we scaled $\langle 
{\cal I}^2 \rangle$ by $C(0)$ and the temperature by $\Gamma$. We 
observe that while the curves for $N \agt 8$ tends to fall onto each 
other, the $N=4$ case shows a marked different behavior.

The limit case of $N=2$ is special: the variance of the pumping response 
function $\Pi_0$ diverges due to the long tails in the distribution 
$\Pi_0$. \cite{Brouwer98} Hence, in the absence of dephasing processes, 
$\langle I_T^2 \rangle$ also diverges. We address this issue in the 
following subsection.

\subsection{Dephasing}  
\label{sec:dephasing-standard-pumping}  
 
The origin of the quantum pumping current is interference. Thus, the 
effect is susceptible to dephasing created by the interaction of 
electrons with phonons, photons, and fluctuations in the 
electromagnetic environment. While a precise, accessible microscopic 
theory of dephasing for open quantum pumps is still lacking, some 
quantitative results can be obtained through phenomenological 
models.\cite{Moskalets01,Cremers02} In particular, the voltage probe 
model of B\"uttiker \cite{Buttiker86} provides a simple way of 
introducing dephasing by adding a third lead to the QD which neither 
inject nor drain a net current. Electrons can move in and out of the 
QD through this lead and loose phase coherence in the process. The 
amount of dephasing can then be tuned by changing the characteristics 
of the third lead, such as coupling constant $p$ and number of 
channels $N_\phi$. It is customary to use a single parameter $P_\phi = 
p N_\phi$ to parametrize the dephasing. We take $N_\phi \gg 1$ and $p 
\ll 1$, while keeping $P_\phi$ constant. 
 
Here we adopt the formulation of Ref. \onlinecite{Cremers02} and break 
up the contributions to the pumping current into two parts, 
\be  
I_0^\phi = \frac{\omega e}{2\pi} \int_{\cal{A}} dX_1 dX_2\, \left( 
\Pi^{\text{dir}}_0 + \Pi^{\text{rec}}_0 \right). 
\ee 
The separation into two components is convenient because 
$\Pi^{\text{dir}}_0$ becomes $\Pi_0^L$ [Eq. (\ref{eq:defPi0})] as the 
coupling between the third lead and the QD goes to zero (see below), 
while $\Pi^{\text{rec}}_0$ disappears in the same limit. The latter 
represents the contribution to the pumping current coming from a 
voltage rectification effect (See Sec. \ref{sec:rectification}).

\begin{figure} 
\includegraphics[width=9cm]{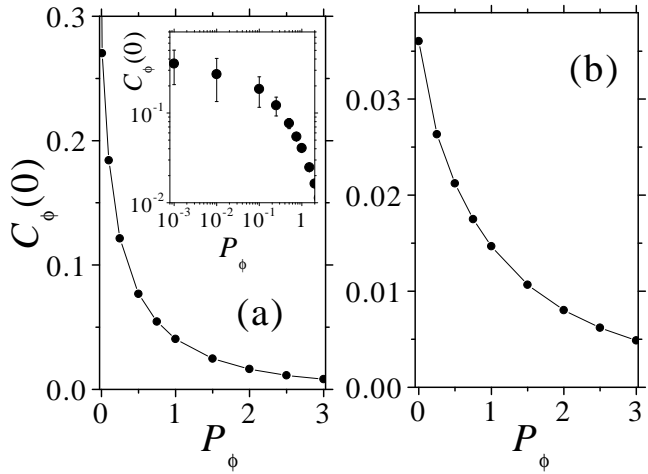}   
\caption{Pumping response variance $C_\phi(0)={\rm var} (\Pi_\phi)$ as 
a function of $P_\phi$. Solid circles are the result of our numerical 
simulation (solid lines serve as a guide to the eye). (a) $N=2$ case. 
Inset: Same as before, but magnifying the small $P_\phi$ range 
(statistical standard deviations shown). (b) $N=4$ case.} 
\label{fig:uvarpumpphi-N2N4}   
\end{figure}   

Using that $\Pi_0^L +\Pi_0^R +\Pi_0^\phi = 0$ to enforce current 
conservation (where $\phi$ denotes the third lead), one finds the 
relation \cite{Cremers02} 
\be  
\label{eq:iphidir} 
\Pi^{\text{dir}}_0 = \frac{G_{R\phi}}{(G_{L\phi} + G_{R\phi})} 
              \,\Pi_0^L - \frac{G_{L\phi}}{(G_{L\phi} + G_{R\phi})} 
              \,\Pi_0^R, 
\ee 
where $G_{L\phi}$ and $G_{R\phi}$ are the conductances of the QD 
between the third lead and the left or right leads, 
respectively. Through steps analogous to those taken to write 
Eqs. (\ref{eq:dQ}) to (\ref{eq:defPi0}), one finds that the 
rectification current is given by 
\ba   
\label{eq:iphirect}   
\Pi^{\text{rect}}_0 &\!\!=&\!\! \frac{1}{4\pi} \sum_\beta\sum_{\alpha 
   \in \phi} \text{Im}\!\left(S^*_{\alpha\beta}\frac{\partial 
   S_{\alpha\beta}} {\partial X_2}\right) \frac{\partial}{\partial 
   X_1} \frac{G_{L\phi} - G_{R\phi}} {G_{L\phi} + G_{R\phi}} \nonumber 
   \\ &&\!\!\!\!\!\!-\frac{1}{4\pi} \sum_\beta\sum_{\alpha \in 
   \phi}\text{Im} \!\left(S^*_{\alpha\beta}\frac{\partial 
   S_{\alpha\beta}} {\partial X_1}\right) \frac{\partial}{\partial 
   X_2} \frac{G_{L\phi} - G_{R\phi}} {G_{L\phi} + 
   G_{R\phi}}. \nonumber \\ 
\ea  
Below, we study the statistical fluctuations of $\Pi_\phi \equiv 
   \Pi^{\text{dir}}_0 + \Pi^{\text{rec}}_0 $.

\begin{figure} 
\vskip-0.45cm   
\includegraphics[width=9cm]{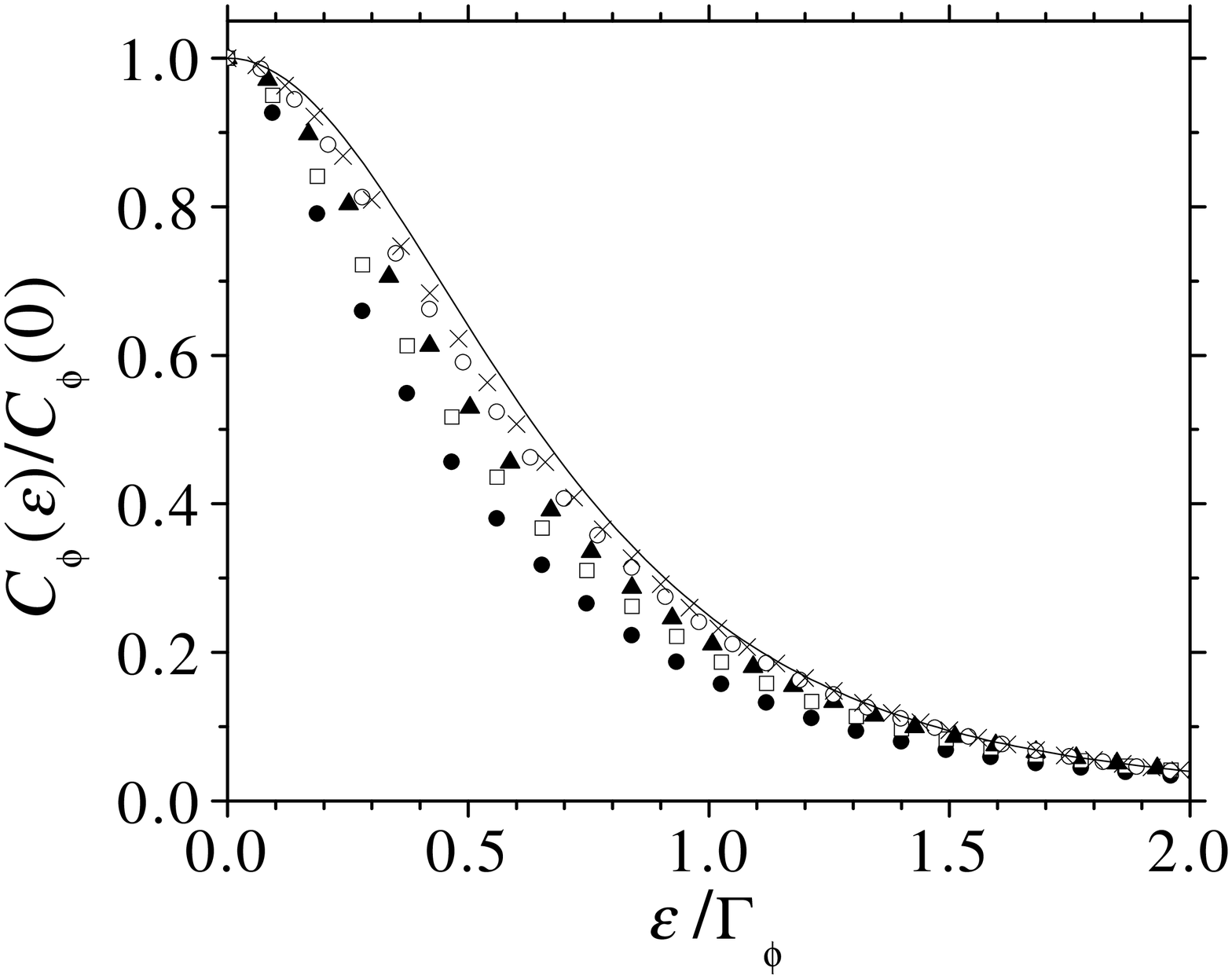}   
\caption{Normalized pumping response $\Pi_\phi$ autocorrelation 
function $C_\phi(\varepsilon)/ C_\phi(0)$ versus 
$\varepsilon/\Gamma_\phi$ for $N_R=N_L=2$. The solid circles, squares, 
triangles, empty circles, and crosses stand for $P_\phi=0.25$, $0.5$, 
$1.0$, $2.0$, and $3.0$, respectively. The solid line stands for Eq.\ 
(\ref{eq:sccorrfunction}).} 
\label{fig:ucorphi_scaled}   
\end{figure}   

In Fig.\ \ref{fig:uvarpumpphi-N2N4} we show the effect of dephasing in 
the pumping response variance, $C_\phi(0) \equiv {\rm var}(\Pi_\phi)$, 
as a function of the dephasing parameter $P_\phi$ for $N=2$ and 
4. When $N=2$, the long tails in the distribution of $\Pi_0$ make the 
numerical assessment of var$(\Pi_\phi)$ increasingly difficult as 
$P_\phi$ goes to zero (see inset of Fig.\ 
\ref{fig:uvarpumpphi-N2N4}a). To investigate this problem, we 
generated $10^6$ realizations of the $S$ matrix for a range of 
energies $E$ and $P_\phi < 0.1$. We found that the standard deviation 
increases with the number of realizations; moreover, large values of 
$|\Pi_\phi|$ were accompanied by abrupt fluctuations of $\Pi_\phi$ 
with $E$. Upon shrinking the energy steps in our numerical 
calculations, we found that the number of large fluctuations 
decreased, but their amplitudes increased. This scaling procedure was 
computationally very costly; yet, despite the somewhat poor 
statistics, our results suggest a fractal behavior of $\Pi_\phi$ as a 
function of energy, typical of power-law 
distributions.\cite{Huckestein00} Therefore, our simulations strongly 
support that $C_\phi(0)$ diverges for $N=2$ as $P_\phi\rightarrow 0$, 
in agreement with Ref. \onlinecite{Brouwer98}. The apparent fractal 
nature of $\Pi_\phi(E)$ in this case indicates that exchanging the 
order between thermal and ensemble averages would likely not render a 
finite value for $\langle I_T^2 \rangle$ either. 
 
For larger values of the dephasing parameter, say $P_\phi \agt 0.1$, 
we obtained a converged value var$(\Pi_\phi)$ (small standard 
deviation) with $10^5$ realizations, despite of the very large values 
of the higher moments. For $P_\phi > 1$, convergence was attained 
already with $10^4$ realizations. We observed (not shown here) that 
our numerical simulations agree with the analytical results for 
$C_\phi(0)$ presented in Ref. \onlinecite{Cremers02}. 
 
In practice, at very low temperatures, var$(\Pi_\phi)$ is extremely 
sensitive to small changes in $P_\phi$. In this case, although in 
principle possible, it is very difficult to give a quantitative 
numerical description of both dephasing and temperature effects. We do 
not pursue this path here and direct the discussion instead to $N>2$, 
where certainly no divergences occur.

\begin{figure}   
\includegraphics[width=9.0cm]{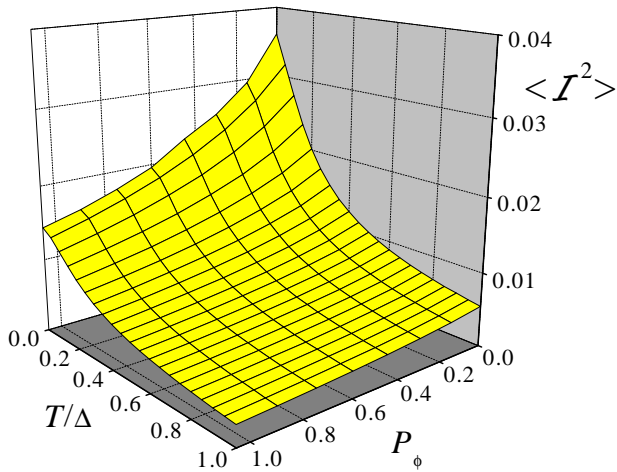}
\caption{$\langle {\cal I}^2 \rangle$ for $N=4$ as a function of the 
temperature and the dephasing strength.} 
\label{fig:varpumpdephasing}   
\end{figure}   

A very strong dephasing dependence is absent for $N=4$. Here, our 
simulations show that $C_\phi(0)$ remains finite as $P_\phi$ 
approaches zero, as depicted by Fig.\ \ref{fig:uvarpumpphi-N2N4}b and 
in quantitative agreement with Eq.\ (\ref{eq:varPi0}). In Fig.\ 
\ref{fig:ucorphi_scaled} we show the effect dephasing in the 
dimensionless pumping energy autocorrelation function. The correlation 
function $C_\phi(\varepsilon)$ is defined as in Eqs.\ (\ref{eq:defD}) 
and (\ref{eq:defC}). We fix $N_R=N_L=2$ and vary the dephasing 
parameter $P_\phi$. The results represent an average over $10^4$ 
realizations. We find empirically that the energy correlation length 
scales as 
\be 
\label{eq:Gammaphi} 
\Gamma_{\phi} = \frac{\Delta}{2 \pi}(N + P_\phi) \,. 
\ee 
Similarly to the case without dephasing, as $P_\phi$ increases one 
quickly reaches the semiclassical regime, characterized by the 
universal correlation function given by Eq. (\ref{eq:sccorrfunction}). 
We verified (not shown here) that our simulations for $C_\phi(0)$ 
coincide with the analytical results obtained in 
Ref. \onlinecite{Cremers02}, as expected. 
 
At this point we can compare the relative role of dephasing and 
temperature for the pumping current. This is done in Fig.\ 
\ref{fig:varpumpdephasing}, where we display $\langle {\cal I}^2 
\rangle \equiv \langle {I_T}^2 \rangle/(e\omega {\cal A}/2\pi)^2$ as a 
function of $T$ and $P_\phi$.  For $N_R=N_L\ge 2$ we find that, at low 
temperatures, where typical QDs have for instance $T/\Delta \approx 
0.5$ and $P_\phi\approx 0.5$,\cite{Huibers98} temperature plays a 
significantly more important role than dephasing. The exception is the 
remarkable case of $N_R=N_L=1$ where, in the absence of dephasing, the 
variance of the pumping current diverges.

\section{Rectified currents} 
\label{sec:rectification} 
 
In AlGaAs/GaAs quantum dots formed by lateral electrostatic gates, the 
rectified component of the dc current tends to dominate over the 
quantum pumping one at low driving frequencies. This has been recently 
verified in an experiment where the symmetry properties of induced 
currents with respect to an external perpendicular magnetic field were 
studied.\cite{dicarlo03} It was found that while pure quantum pumping 
currents should be asymmetric with respect to field 
inversion,\cite{Shutenko00} the actual measured current showed a 
strong even symmetry, characteristic of rectification.\cite{Brouwer01} 
It is therefore important to characterize the statistical fluctuations 
of the rectified current as well. 
 
Rectification can arise from two different sources. Firstly, as 
mentioned in Sec. \ref{sec:dephasing-standard-pumping}, it appears 
when inelastic processes take place inside the QD, leading to 
dephasing. When one models the dephasing process by allowing carriers 
to move in and out of the QD incoherently through a third lead, an 
additional reservoir with time-dependent chemical potential, 
$\mu_\phi(t)$, is required.\cite{Moskalets01} Since the time 
modulation of the QD shape causes the conductances to vary in time, an 
incoherent dc current can flow between left and right reservoirs, as 
long as $G_{L\phi}$ and $G_{R\phi}$ are nonzero. This effect is 
intrinsic to the QD and has therefore been included in the expression 
for the quantum pumping current. We remark that the rectified current 
in this case is also asymmetric with respect to magnetic field 
inversion.\cite{Cremers02} For an open QD, this contribution to the 
current vanishes as the dephasing strength decreases. 
 
\begin{figure}   
\includegraphics[width=6.0cm]{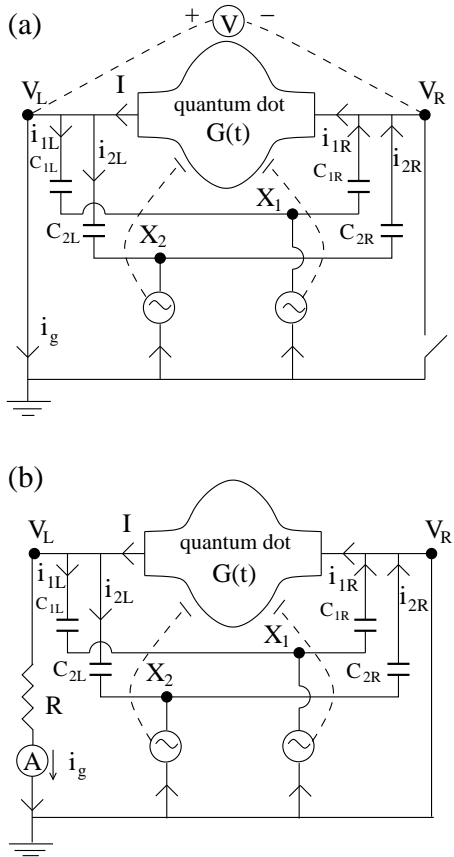} 
\caption{Equivalent circuits for the quantum pumping measurements: (a) 
voltage setup; (b) current setup. The electrodes are coupled to the 
right (left) reservoir through capacitances $C_{1R}$ and $C_{2R}$ 
($C_{1L}$ and $C_{2L}$), with displacement currents $i_{1R}$ and 
$i_{2R}$ ($i_{1L}$ and $i_{2L}$). In (a) the reservoir-dot-reservoir 
circuit is open, while in (b) both reservoirs are grounded. $R$ 
accounts for the circuit and the current meter internal 
resistance. $X_1$ and $X_2$ are the ac voltages in the electrodes and 
$G(t)$ is the quantum dot conductance.} 
\label{fig:rectif}   
\end{figure}

Secondly, rectification also appears due to the capacitive coupling 
between the shape-deforming electrodes and the left and right 
reservoirs.\cite{Brouwer01,switkes99phd} The displacement currents and 
the conductance of the QD oscillate with the same frequency, producing 
a net dc charge current between the reservoirs. The exact way by which 
the dc current is induced depends on the particular measurement 
setup. To illustrate this point, in Fig. \ref{fig:rectif} we show 
schematically the two equivalent circuits, namely, for (a) voltage or 
(b) current measurements.\cite{switkes99phd} Let us briefly derive 
expressions for the QD voltage and current for these setups, expanding 
the discussion found in Ref. \onlinecite{Brouwer01}.

\noindent 
(a) {\it Voltage setup}: in this case the reservoir-QD-reservoir loop 
is open. Calling $V = V_R - V_L$, with $V_L=0$, we find that 
\be  
I = C_{2R} \frac{d}{dt} \left( X_2 - V \right) + C_{1R} 
\frac{d}{dt} \left( X_1 - V \right). 
\ee 
Since $I = G\, V$ as well, we arrive at 
\be  
V = \frac{1}{G} \left[ C_{2R} \frac{d}{dt} \left( X_2 - V \right) 
+ C_{1R} \frac{d}{dt} \left( X_1 - V \right) \right]. 
\ee 
For sufficiently adiabatic pumping (MHz) and small capacitive 
couplings (pF), we can have $V \ll X_{1,2}$, given that $G^{-1} \leq 
26$ K$\Omega$. As a result,\cite{Brouwer01} 
\be 
\label{eq:rectV} 
V \approx \frac{1}{G} \left( C_{2R} \frac{dX_2}{dt} + C_{1R} 
\frac{dX_1}{dt} \right). 
\ee 
(b) {\it Current setup}: in this case both reservoirs are grounded and 
$V_R=0$. Since $i_g = I - i_{1L} - i_{2L}$, we find 
\be 
I = \frac{V}{R} + C_{2L} \frac{d}{dt} \left( V - X_2 \right) + C_{1L} 
\frac{d}{dt} \left( V - X_1 \right). 
\ee 
Using $I = G\, V$, we get 
\be  
\label{eq:rectI} 
I = \frac{RG}{1 - RG} \left[ C_{2L} \frac{d}{dt} \left( X_1 - 
V \right) + C_{1L} \frac{d}{dt} \left( X_2 - V \right) \right]. 
\ee 
Here we can also neglect $V$ with respect to $X_{1,2}$ on the 
right-hand side of Eq. (\ref{eq:rectI}). Moreover, in the MHz range, 
where the pumping experiments are carried out, the impedance of the 
current meter should be smaller than the quantum dot resistance, namely, 
$RG \alt 1$.\cite{marcuscomm} Thus, we finally obtain\cite{Brouwer01} 
\be 
\label{eq:rectI1} 
I \approx RG \left( C_{2L} \frac{dX_1}{dt} + C_{1L} 
\frac{dX_2}{dt} \right). 
\ee

Focusing now in the current setup [Eq. (\ref{eq:rectI1})], it is 
straightforward to show that an expression for the rectified current 
similar to Eq. (\ref{eq:defI0}) exists, with the kernel 
\be  
\Pi^{\text{cap}}_0 = RC_2 \frac{\partial G}{\partial X_1} - RC_1  
\frac{\partial G}{\partial X_2}, 
\ee  
where we call $\partial G/\partial X_i$ the parametric conductance 
velocity with respect to $X_i$. The statistical fluctuations of this 
quantity were investigated in Ref. \onlinecite{Brouwer97}, where the 
distributions of $\partial G/\partial X$ were presented for the $N=2$ 
case. Recall that the linear conductance $G$ is given by the Landauer 
formula 
\be  
G = \frac{e^2}{h}\, g_0 , 
\ee 
with the dimensionless conductance at zero temperature defined as 
\be  
g_0 = \sum_{{\alpha\in L}\atop{\beta\in R}} \left| S_{\alpha\beta} 
\right|^2. 
\ee 

It is difficult to estimate the amplitude of the rectified current 
derived from $\Pi^{\text{cap}}_0$, since it depends on $R$ which is 
unknown. However, the symmetries and statistical properties can be 
determined and should differ from those of the pure quantum pumping 
current component. For instance, notice that $\Pi^{\text{cap}}_0$ 
depends directly on $\partial g_0/\partial X$. Since the conductance 
is symmetric with respect to field inversion, and so will be the 
rectified component of the current in 
Eq. (\ref{eq:rectI1}).\cite{Brouwer01} Let us focus now on the 
statistical properties of $\Pi^{\text{cap}}_0$ and on its temperature 
and dephasing dependences. We argue that thermal smearing plays a key 
role in suppressing fluctuations of the rectified current as well. In 
App. \ref{sec:appP(dGdX)}, we study a very closely related issue and 
show how thermal smearing explains the discrepancy between the 
recently experimental measured parametric conductance velocity 
distribution \cite{Huibers98} and the analytical predictions at zero 
temperature.\cite{Brouwer97}

\begin{figure}   
\includegraphics[width=9cm]{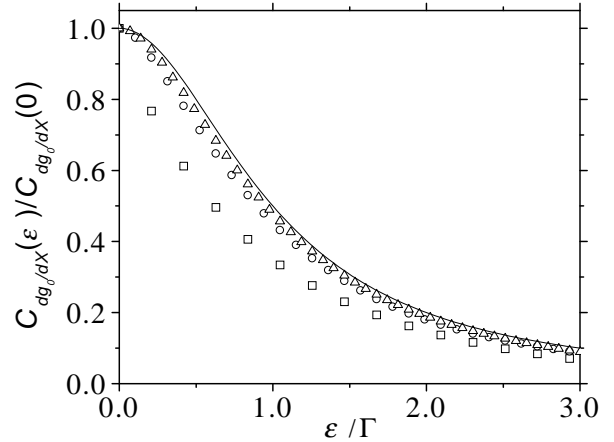}   
\caption{Parametric conductance velocities autocorrelation function 
$C_{\partial g_0/\partial X}(\varepsilon)$ as a function of 
$\varepsilon/\Gamma$. The solid line stands for Eq.\ (\ref{eq:CdGdX}), 
while squares, circles, and triangles stand for $N_L=N_R=1$, $2$, and 
$3$ respectively.} 
\label{fig:cappumpcor}    
\end{figure}   

A simple inspection of Eqs. (\ref{eq:S}) and (\ref{eq:dSdX}) shows 
that $\langle\partial g_0/ \partial X_i\rangle = 0$, since it is 
proportional to $\langle\partial H/ \partial X_i\rangle$. Hence, 
$\langle \Pi^{\text{cap}}_0\rangle = 0$. The variance of 
$\Pi^{\text{cap}}_0$, on the other hand, is given by 
\ba 
& & \text{var}(\Pi^{\text{cap}}_0) = \left( \frac{Re^2}{h}\right)^2 
\left[ C_2^2 \left\langle \left( \frac{\partial g_0}{\partial 
X_1}\right)^2 \right\rangle + \right.  \nonumber \\ && \left. \!\!\! 
C_1^2 \left\langle \left(\frac{\partial g_0}{\partial X_2}\right)^2 
\right\rangle - 2 C_1C_2 \left\langle \frac{\partial g_0}{\partial 
X_1} \frac{\partial g_0}{\partial X_2} \right\rangle \right]. 
\ea 
For chaotic systems, provided the two parametric perturbations 
$X_1(t)$ and $X_2(t)$ have the same amplitude, we expect that 
\be  
\left\langle \left(\frac{\partial g_0}{\partial  
X_1}\right)^2\right\rangle = \left\langle \left(\frac{\partial  
g_0}{\partial X_2}\right)^2\right\rangle \equiv \left\langle  
\left(\frac{\partial g_0}{\partial X}\right)^2\right\rangle \,.  
\ee  
If the periodic perturbations $X_1$ and $X_2$ are acting at locations 
far (by several Fermi wavelengths) from each other, it justifiable to 
assume that the parametric conductance velocities with respect to 
different $X_i$ are uncorrelated, $\langle (\partial g_0/\partial 
X_1)(\partial g_0/\partial X_2)\rangle =0$, leading to $\text{var} \, 
(\Pi^{\text{cap}}_0)= R^2 ( C_1^2 + C_2^2 ) \langle (\partial 
g_0/\partial X)^2 \rangle$. This assumption leads to an analytical 
expression for the zero-temperature rectified current variance, 
var($i^{\text{rect}}$), in analogy with Eq. (\ref{eq:defI0}), since 
\be    
\label{eq:var(dGdX)}   
\left\langle \left( \frac{\partial g_0}{\partial X} \right)^2  
\right\rangle =  
\frac{8 \, N_L^2 N_R^2}{N \, (N^2-1)^2} .  
\ee  
[Details on the derivation of Eq.\ (\ref{eq:var(dGdX)}) are presented 
in Appendix \ref{sec:appDeriva_var(dGdX)}.] In order to compute the 
temperature dependence of var($i^{\text{rect}}$) we need first to 
calculate the parametric conductance velocity energy autocorrelation 
function, namely, 
\be C_{\partial g_0/\partial X} (\varepsilon) =  
\left\langle \frac{\partial g_0}{\partial X} 
\!\left(E+\frac{\varepsilon}{2}\right) \frac{\partial  
g_0}{\partial X}\!\left(E -\frac{\varepsilon}{2}\right)\right\rangle  
\ee  
(recall that $\langle \partial g_0/ \partial X\rangle = 0$). In 
Fig. \ref{fig:cappumpcor} we show the results of our simulations for 
various values of $N$. Again, as $N$ increases $C_{\partial 
g_0/\partial X} (\varepsilon)$ very rapidly converges to 
\be   
\label{eq:CdGdX}   
C_{\partial g_0/\partial X}(\varepsilon) =  
\frac{C_{\partial g_0/\partial X}(0)} {1 + (\varepsilon/\Gamma)^2}, 
\ee  
where $C_{\partial g_0/\partial X}(0)$ is given by Eq. 
(\ref{eq:var(dGdX)}), while deviations are quite large for small $N$. 
As in Sec.\ \ref{sec:smallN}, the thermal fluctuations are enhanced in 
the small-channel case. 
 
For completeness, let us discuss now the opposite, correlated case, 
where $\langle (\partial g_0/ \partial X_1) (\partial g_0/ \partial 
X_2)\rangle = (\partial g_0/ \partial X)^2$. Now $\text{var} \, 
(\Pi^{\text{cap}}_0) = R^2 ( C_1 - C_2)^2 \langle (\partial 
g_0/\partial X)^2 \rangle$, making and the effect of rectification 
very small when the capacitive coupling is close to symmetric ($C_1 
\approx C_2$). However, it is unlikely that such condition is 
satisfied in real experimental setups.

\begin{figure}   
\includegraphics[width=9cm]{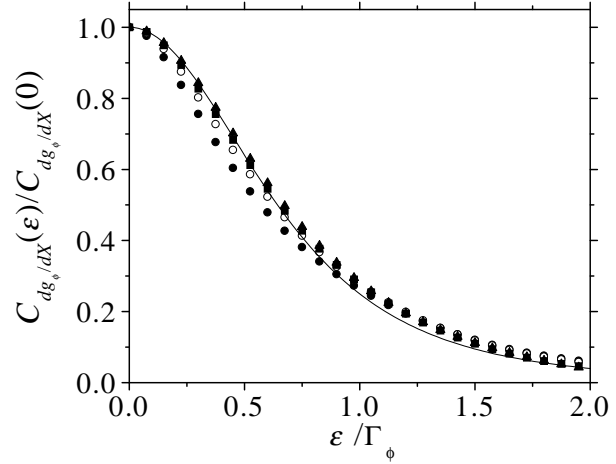}
\caption{Parametric conductance velocities autocorrelation function 
$C_{\partial g_\phi/\partial X}(\varepsilon)$ as a function of 
$\varepsilon/\Gamma_\phi$ in the presence of dephasing for 
$N_L=N_R=1$. The solid line stands for Eq.\ (\ref{eq:CdGdX}); squares, 
circles, triangles, and inverted triangles stand for $P_\phi=0.5$, 
$1$, $2$, and $4$, respectively.} 
\label{fig:cappumpcor2}    
\end{figure}   


Dephasing effects can also be included in $\Pi^{\text{cap}}_0$ 
phenomenologically by using the voltage probe model in its original 
form.\cite{Buttiker86} For that purpose, the dimensionless conductance 
is replaced by 
\be  
g = g_0 + \frac{g_{L\phi}g_{\phi R}}{g_{L\phi}+g_{R\phi}}. 
\ee  
We have calculated the distribution of $\partial g_\phi /\partial X$ 
for several values of $N$ and the dephasing parameter $P_\phi$, as 
introduced in Sec. \ref{sec:dephasing-standard-pumping}. The 
conductance velocity autocorrelation function scales in the same way 
as in the case without dephasing, provided we replace $\Gamma$ by 
$\Gamma_\phi$. Results for the case $N=2$ are shown in 
Fig. \ref{fig:cappumpcor2}. The dependence with $P_\phi$ is weaker 
than in the pure quantum pumping case (compare with 
Fig. \ref{fig:ucorphi_scaled}).

\section{Spin currents due to pumping and rectification}   
\label{sec:spin}   
 
It was recently pointed out that it is possible to pump a spin current 
without any net flow of charge through a QD.\cite{Mucciolo02} The 
basic idea is to use a parallel magnetic field to Zeeman split 
resonant states around the Fermi energy. By doing so and tuning the QD 
shape, one can find a situation where up and down spin components of 
the pumping current flow in opposite directions. The mechanism by 
which charge is pumped can be either the ideal quantum pumping of 
Sec. \ref{sec:theory} or the rectification due to the capacitive 
coupling between pumping electrodes and reservoirs, as described in 
Sec. \ref{sec:rectification}. A recent experiment has observed the 
effect.\cite{Watson03} 
 
An important practical question is by how much is the spin current 
attenuated by dephasing in the orbital (charge) sector of the 
wavefunctions. Furthermore, it is also interesting to know the 
amplitude of the effect when the spin pumping current is caused by 
rectification alone and the number of propagating channels in the 
leads is not large (both cases had not being considered in the 
analysis of Ref. \onlinecite{Mucciolo02}) 
 
The dependence of the spin current polarization amplitude on the 
applied parallel field and temperature can be estimated from the 
correlator $\langle I_\uparrow\, I_\downarrow \rangle$. Here 
$I_{\uparrow,\downarrow} = I_T(\mu \pm E_Z/2)$, where $E_Z = g^\ast 
\mu_B\, B_\parallel/2$ is the Zeeman energy. Using the relations 
presented in Sec. \ref{sec:theory}, it is straightforward to show that 
the correlator can be written as 
\be   
\label{eq:avIupIdown}   
\langle I_\uparrow I_\downarrow \rangle = T^2 \int_{-\infty}^{\infty} 
d \varepsilon \frac{d}{dT} \left[ 2 T \,\mbox{sinh}\! \left(  
\frac{\varepsilon}{2T} \right) \right]^{-2} C_0(\varepsilon + E_Z). 
\ee  
This expression is valid regardless of the nature of the pumping 
mechanism.

\begin{figure}   
\includegraphics[width=9cm]{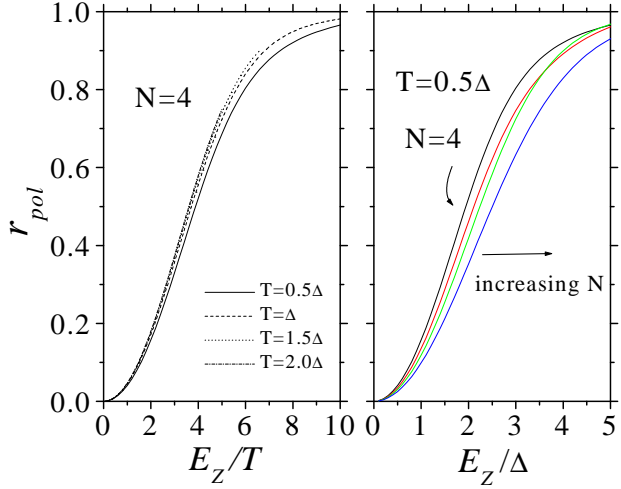}   
\caption{Spin pumping current polarization coefficient as a function 
of Zeeman energy $E_Z$ (parallel magnetic field). (a) Curves for 
different temperatures and a fixed number of channels ($N_R = N_L 
=2$). (b) Dependence on the number of channels for a fixed 
temperature.} 
\label{fig:rpolkT05} 
\end{figure}   

As discussed in Sec. \ref{sec:semiclasSmatcor}, in the case of pure 
quantum pumping, Eq. (\ref{eq:sccorrfunction}) provides a 
semiclassical approximation to the energy correlator in the absence of 
dephasing and when $N \gg 1$. Using that expression and carrying out 
simple manipulations, we find that Eq. (\ref{eq:avIupIdown}) becomes 
identical to Eq. (3) of Ref. \onlinecite{Mucciolo02}, namely, 
\ba 
\label{eq:spinpumpsemiclass} 
\langle I_\uparrow I_\downarrow \rangle & = & C_0(0) \Gamma 
\int_0^\infty d\tau\, (1 + \Gamma \tau) \,e^{-\Gamma \tau} \left[ 
\frac{\pi T\tau}{\sinh(\pi T\tau)} \right]^2 \nonumber \\ & & \times 
\cos(E_z \tau), 
\ea 
The latter was originally derived in Ref. \onlinecite{Mucciolo02} from 
the linear-response limit of the large-$N$ diagrammatic formalism 
developed in Ref. \onlinecite{Vavilov01b}. Notice that when dephasing 
is present, $\Gamma$ has to be replace by $\Gamma_\phi$, as defined in 
Eq. (\ref{eq:Gammaphi}), but the functional dependence on $T$ and 
$E_Z$ remains the same.

In the small-$N$ limit, we do not have a closed analytical expression 
for the energy correlator. In this case, the dependence of $\langle 
I_\uparrow I_\downarrow \rangle$ on temperature and magnetic field 
amplitude can only be obtained numerically. Our results for this case 
are presented in Fig. \ref{fig:rpolkT05}, together with curves derived 
from Eq. (\ref{eq:spinpumpsemiclass}). We have opted for plotting the 
spin current amplitude in terms of the spin polarization coefficient, 
defined as 
\be  
r_{\text{pol}} = \frac{ 1 - \langle I_\uparrow I_\downarrow \rangle / 
\langle I_\uparrow^2 \rangle }{ 1 + \langle I_\uparrow I_\downarrow 
\rangle / \langle I_\uparrow^2 \rangle }. 
\ee 
Notice the rather weak temperature dependence after rescaling the 
Zeeman energy. In contrast, the dependence on the number of channels 
is much more pronounced; consequently, a strong dependence on the 
dephasing parameter $P_\phi$ also occurs. Thus, the same restrictions 
to orbital (charge) quantum pumping due to decoherence by the 
environment apply to the spin case. 
 
\begin{figure}   
\includegraphics[width=9cm]{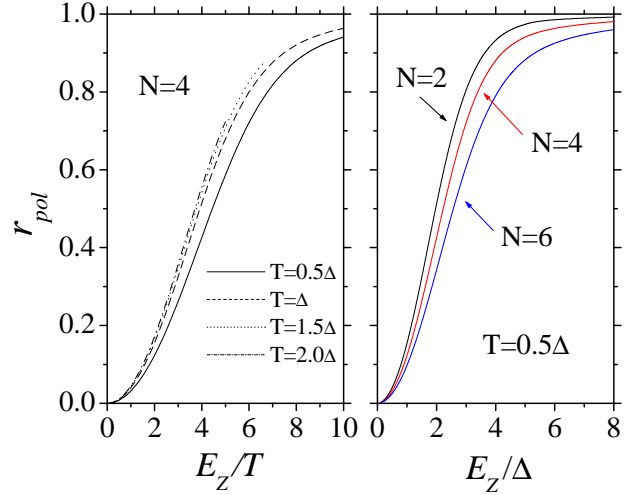} 
\caption{Same as in Fig. \ref{fig:rpolkT05}, but for 
rectification-induced spin pumping currents.} 
\label{fig:capspinpump} 
\end{figure}

Equation (\ref{eq:avIupIdown}) is also valid when rectification is 
present and dominates the pumping mechanism. In this case, in the 
large-$N$ limit, the energy correlator of Eq. (\ref{eq:CdGdX}) can be 
used, leading to an expression very similar to 
Eq. (\ref{eq:spinpumpsemiclass}), namely, 
\be  
\langle I_\uparrow I_\downarrow \rangle = 2\, C_0(0) \Gamma 
\int_0^\infty d\tau\,e^{-\Gamma \tau} \left[ \frac{\pi 
T\tau}{\sinh(\pi T\tau)} \right]^2 \cos(E_z \tau). 
\ee 
As in the dissipationless case, here we have to rely on numerical 
calculations to obtain the correlator in the small-$N$ limit (see 
Sec. \ref{sec:rectification}). Comparative results are shown in 
Fig. (\ref{fig:capspinpump}) as a function of Zeeman energy (magnetic 
field) and for different temperatures and number of channels. The 
curves are very similar to those of Fig. \ref{fig:rpolkT05}. The 
dependence on temperature is just slightly more pronounced for 
capacitive pumping. Considering that the energy correlators for the 
two pumping mechanisms are very similar (a Lorentzian square for pure 
quantum as opposed to a simple Lorentzian for the capacitive case, in 
the semiclassical limit), this result comes as now surprise. We 
conclude that temperature and dephasing dependences alone are not 
strong indicators of the nature of the pumping mechanism, both for the 
charge and the spin cases.

\section{Conclusions} 
\label{sec:conclusions} 
 
In this work we presented numerical (small-$N$) and semiclassical
(large-$N$) calculations of statistical measures of charge and spin
pumping currents in open chaotic quantum dots. Both pure quantum and
rectification pumping mechanisms were considered when time-reversal
symmetry is broken and the regime is adiabatic. We paid particular
attention to the dependences on temperature $T$, number of propagating
channels $N$, and dephasing. We were able to draw several conclusions
from our results.
 
Our initial motivation was to propose measuring the damping of pumping
current mesoscopic fluctuations as a way to estimate the amount of
dephasing present in the experimental setups. However, we found that,
for realistic conditions, thermal smearing tends to be a more
important effect than dephasing. Lower temperatures are required
($T\ll \Delta$) for dephasing to become quantitative and qualitative
distinguishable.
 
Perhaps one situation where dephasing does have a stronger impact than
thermal smearing is when $N=2$, namely, when there is one single
propagating channel in each lead. In this case, dephasing cuts off the
divergence of the zero-temperature pure pumping current
variance. Taking the thermal average before the ensemble average, on
the other hand, does not seem to yield a finite variance when
dephasing is absent. In fact, our numerical simulations suggest that
the zero-temperature pumping current response function, $\Pi_0$, has a
fractal behavior for $N=2$. A long tail in its probability
distribution was observed. Events belonging to this tail were
connected to high but isolated peaks in $\Pi_0$ as the energy is
varied. We observed that such peaks proliferate as the energy
increment decreases. Albeit not yet accessible to experimental
investigation, a more quantitative understanding of the suppression of
this fractal pattern by dephasing is necessary.
 
We also derived an expression for the pumping response autocorrelation
function using the semiclassical approximation in the large-$N$
limit. Our result coincides in the leading order in powers of $1/N$
with that reported in Ref. \onlinecite{Shutenko00}, which was obtained
by the diagrammatic technique. The simplicity of the semiclassical
derivation relating the pumping current to the instability of the
underlying classical orbits provides additional insight to the
physical process. In addition, our simulations show that the
semiclassical limit is very quickly attained. We also calculated the
pumping response variance using the $S$-matrix maximum entropy
approach for pure quantum and rectification pumping. The results are
valid for any number of channels and are both consistent with the
numerical simulations and with the results presented in
Ref. \onlinecite{Cremers02}.

For the rectification case, we found that the expressions relating the
pumping currents and voltages to the driving perturbations are
somewhat more involved than that proposed in
Ref. \onlinecite{Brouwer01}. Our results only coincide with those of
Ref. \onlinecite{Brouwer01} in the limit of weak capacitive coupling
between plunger gates and leads, and when frequencies are not too high
(so that the internal impedance of the current meter remains lower
than the quantum dot resistance).

Rectification currents induced by capacitive couplding also show
mesoscopic fluctuations. However, dephasing effects are less
pronounced than for pure quantum pumping. Moreover, the variance does
not diverge at zero temperatures for any value of $N\ge 2$.
 
Pure quantum pumping spin currents have the same characteristics of
the charge case. Namely, their variance diverges in the absence of
dephasing. We found that increases in temperature, dephasing, or $N$
suppress spin polarization in a similar (strong) way. New mechanisms
of pure spin currents generations based on pumping have been proposed
recently. Particularly attractive are those based on the spin-orbit
coupling present in two-dimensional electron gases formed in III-V
heterostructures, which do not require the application of large
magnetic fields.\cite{fazio02,sharma03} For the future, we plan to
study how thermal smearing and orbital and spin dephasing affect the
magnitude of these new mechanisms.

\acknowledgments 
   
We thank P. Brouwer, G. Finkelstein, C. Marcus, and B. Reulet for
useful discussions. MMM is supported by CLAF-CNPq. CHL and ERM
acknowledge partial support in Brazil from PRONEX, CNPq, Instituto do
Mil\^enio de Nanoci\^encias, and FAPERJ. CHL thanks CBPF (Brazil) and
the Department of Physics at Duke University for the hospitality. This
work was supported in part by NSF Grant No. DMR-0103003.

\appendix
 
\section{Mean and variance of the pumping response function} 
\label{sec:appDeriva_var(Pi0)} 
 
Here we derive Eq.\ (\ref{eq:varPi0}). It is convenient to parametrize 
the $S$ matrix, its parametric-derivatives, and its energy derivative 
as 
\be  
\label{parSdX1SdX2SdES} 
S = UV, \quad 
\frac{\partial S}{\partial X_j} = i\, U Q_{X_j} V , \quad 
\text{and} \quad \frac{\partial S}{\partial E} = 
\frac{2\pi i}{\Delta} U Q_E V,  
\ee 
respectively, for $j=1,2$. Here, $U$ and $V$ are $N\times N$ unitary 
matrices uniformly distributed over the unitary group and independent 
of $Q_{X_j}$ and $Q_{E}$. In turn, the latter are $N\times N$ 
Hermitian matrices satisfying the joint distribution \cite{Cremers02} 
\ba  
\label{QEQX1QX2dist}  
& & \!\!\!\!\!\!\!\! P(S, Q_E, Q_{X_1}, Q_{X_2}) \propto \left( \det\, 
Q_E \right)^{-9N/2} 
\nonumber \\ & & \times 
\exp \left\{ -\text{tr} \left[ 
Q_E^{-1} + 
\frac{1}{8} \sum_{j=1}^2 
\left( Q_E^{-1} Q_{X_j} \right)^2 \right] \right\}. \quad
\ea 
Also, we parametrize $Q_{X_j}$ as 
\be  
\label{QXjpar} 
Q_{X_j} = {\Psi^{\dagger}}^{-1} K_j \Psi^{-1} , 
\ee 
where $\Psi$ is a complex $N\times N$ matrix such that 
\be  
\label{QEpar} 
Q_E = {\Psi^{\dagger}}^{-1} \Psi^{-1} 
\ee 
and $K_j$ is an Hermitian $N\times N$ matrix whose elements are 
Gaussian distributed with zero mean and variance, 
\be 
\label{gaussvar} 
\langle K_{ab} K_{a'b'} \rangle = 
4 ( \delta_{ab'} \delta_{ba'} + \delta_{aa'} \delta_{bb'} ), 
\ee 
as can be easily seen substituting Eqs. (\ref{QXjpar}) and 
(\ref{QEpar}) into the distribution (\ref{QEQX1QX2dist}). 
 
We substitute Eq. (\ref{parSdX1SdX2SdES}) into Eq. (\ref{eq:defPi0}) 
and average over the matrices $U$, $V$ using the results of 
Ref. \onlinecite{Mello90}. Next, we average over $Q_{X_j}$ using the 
parametrization of Eq. (\ref{QXjpar}) and Eq. (\ref{gaussvar}). As a 
result, we obtain $\langle \Pi_0 \rangle = 0$ and 
\ba 
\label{varPiL-1} 
{\rm var}\,\Pi_0 \!\!& = & \!\! 
-\frac{8 N_L N_R}{N\left( N^2-1 \right)}\left[\sum_{a,b,c=1}^{N}  
\left\langle (Q_E^2 )_{aa} (Q_E^{})_{bb} (Q_E^{})_{cc} \right\rangle\right.  
\nonumber \\  
\!\!& & \!\!-\left. \sum_{a=1}^{N} \left\langle \left( Q_E^4 
\right)_{aa} \right\rangle \right] \,. 
\ea 
Now, we write $Q_E$ in its diagonal form as $Q_E=A{\hat \tau} 
A^{\dagger}$, where ${\hat \tau}$ is the eigenvalue matrix and $A$ is 
a random unitary matrix. Substitution of the diagonal form of $Q_E$ 
into Eq. (\ref{varPiL-1}) gives a result independent of $A$ such that 
the average over $A$ is easily done. After some algebraic 
simplifications, we arrive at 
\ba  
\label{varPiL-2} 
{\rm var} (\Pi_0) & = & \frac{32 N_L N_R}{N+1} 
\left[ \left\langle \tau_1^2 \tau_2^2 \right\rangle + 
2 \left\langle \tau_1^3 \tau_2 \right\rangle \right. \nonumber \\ & & + 
\left. (N - 2) \left\langle \tau_1^2 \tau_2 \tau_3 \right\rangle \right]. 
\ea 
Finally, since the dimensionless scape rates $x_n=1/\tau_n$ are 
distributed according to the generalized Laguerre 
ensemble,\cite{bfbPRL78} 
\be  
\label{GLaguerre} 
P\left(x_1,\ldots, x_{N}\right) \propto \prod_{n<m} \left( x_n - 
x_m \right)^2 \prod_{n} x_n^{N} e^{-x_n}. 
\ee 
The averages appearing in Eq. (\ref{varPiL-2}) can be calculated by 
direct integration. The result is Eq. (\ref{eq:varPi0}).

\begin{figure}   
\includegraphics[width=8cm]{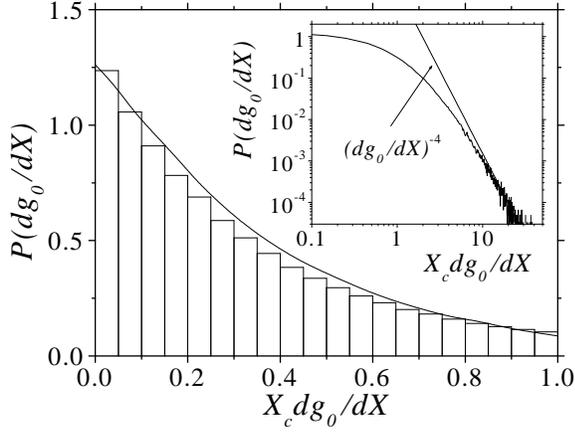}   
\caption{Distribution of parametric conductance velocities $P(\partial 
g_0/\partial X)$ for $N=2$. The histogram is our numerical simulation, 
whereas the solid line stands for the analytical result of 
Ref. \onlinecite{Brouwer97}. In the inset we show the tails of the 
distribution in a log-log scale.} 
\label{fig:cappumphist} 
\end{figure}   

\section{Parametric transmission coefficient velocity variance} 
\label{sec:appDeriva_var(dGdX)} 
 
In this appendix we show how to obtain $\langle (\partial g_0/\partial 
X)^2 \rangle$, as given by Eq.\ (\ref{eq:var(dGdX)}). As in Appendix 
\ref{sec:appDeriva_var(Pi0)}, we find useful to parametrize the $S$ 
matrix as in Eq. (\ref{parSdX1SdX2SdES}). In this case, $Q_X$, $Q_E$, 
and $S$ have the joint distribution \cite{bfbPRL78} 
\ba 
\label{QEQXdist} 
P (S, Q_E, Q_X) & & \!\!\!\! \propto \left( \det\, Q_E \right)^{-4N} 
\nonumber \\ & & \!\!\!\!\!\!\!\!\!\!\!\! \times \exp 
\left\{ \!\!-\text{tr} \left[ Q_E^{-1} 
+ \frac{1}{8} \left( 
Q_E^{-1} Q_X \right)^2 \right]\right\}. 
\ea 
We use Eq. (\ref{parSdX1SdX2SdES}) to express $\langle (\partial 
g_0/\partial X)^2 \rangle$ as a function of $U$, $V$, $Q_X$, and 
$Q_E$. With the help of Ref.\ \onlinecite{Mello90}, we average over 
the matrices $U$ and $V$. Next, using Eqs. (\ref{QXjpar}) and 
(\ref{gaussvar}), we integrate over $Q_X$ to obtain 
\ba \label{vardXT2-QE} \left\langle\left(\frac{\partial g_0}{\partial 
X}\right)^2\right\rangle & = & 8 \left[\frac{N_L N_R}{N\left( N^2-1 
\right)}\right]^2 \text{Re} \left[ - \sum_{a=1}^{N} \left\langle 
\left( Q_E^2 \right)_{aa} \right\rangle \right. \nonumber \\ & + & 
\left. N \sum_{a,b=1}^{N} \left\langle \left( Q_E \right)_{aa} \left( 
Q_E \right)_{bb} \right\rangle \right]. 
\ea 
Again, we write $Q_E$ in its diagonal form as $Q_E=A{\hat \tau} 
A^{\dagger}$ and substitute into Eq. (\ref{vardXT2-QE}). After some 
algebra, we arrive at 
\be 
\left\langle\left(\frac{\partial g_0}{\partial 
X}\right)^2\right\rangle = \frac{8 (N_L N_R)^2}{N( N-1)(N+1)^2} 
\left( \left\langle \tau_1^2 \right\rangle + N \left\langle 
\tau_1\tau_2 \right\rangle \right). 
\ee 
The averages over $\tau_1^2$ and $\tau_1\tau_2$ can be done explicitly 
by direct integration using Eq. (\ref{GLaguerre}). The result is Eq. 
(\ref{eq:var(dGdX)}).

\begin{figure} 
\includegraphics[width=8cm]{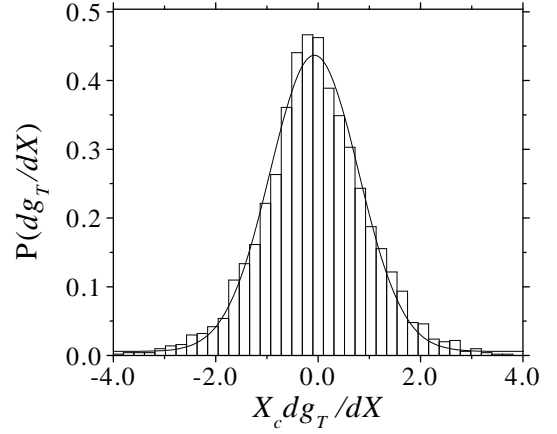}   
\caption{Distribution of parametric conductance velocities $P(\partial 
g_T/\partial X)$, $N_R=N_L=1$, and $T=0.6\Delta$. The solid line 
indicates the best Gaussian fit.} 
\label{fig:Pdgdx_T}   
\end{figure} 

\section{Parametric conductance velocity distributions}   
\label{sec:appP(dGdX)}   
   
This appendix serves to remedy the discrepancy between the theoretical 
\cite{Brouwer97} and experimental \cite{Huibers98} distributions of 
parametric dimensionless velocities $\partial g_0/\partial X$. This 
discussion gives further support to the statistical theory employed in 
this paper, and provides a further illustration of the important role 
of temperature in the statistical fluctuations of conductance-related 
quantities. In addition, we illustrate the accuracy of the simulations 
by comparing our results to the theoretical $P(\partial g_0/\partial 
X)$. 
 
For $N=2$ the distribution of $\partial g_0/\partial X$ is known for 
all symmetry classes.\cite{Brouwer97} It shows a singularity at zero 
derivative: a logarithmic divergence in the presence of time-reversal 
symmetry ($\beta=1$) and a cusp in the absence of that symmetry 
($\beta=2$). The tails of $P(\partial g_0/\partial X)$ are algebraic 
and follow 
\be  
P(\partial g_0/\partial X) \propto (\partial g_0/\partial 
X)^{-\beta-2}. 
\ee 
In Fig. \ref{fig:cappumphist} we show a comparison between our 
numerical results and the analytical distributions of $\partial 
g_0/\partial X$. 

By numerically convoluting $g_0(\epsilon)$ with the thermal 
distribution, we obtain $g_T$. The histogram is show in 
Fig. \ref{fig:Pdgdx_T} for $T/\Delta = 0.6$, corresponding to the 
experimental situation in Ref. \onlinecite{Huibers98}. Notice that 
even at such low temperatures, $T$ is slightly larger than the 
$\partial g_0/\partial X$ energy autocorrelation length, suppressing 
large fluctuations of $g_0$ at zero temperature and favoring a 
Gaussian distribution, as observed in the experiment. 
 
The quality of this result suggests that a direct comparison between 
experimental and theoretical values of var($\partial g_T/\partial X)$ 
can be used to set the scale of the $X$ parameter. The distribution of 
rectified currents $P(i^{\text{rect}})$ can be obtained 
straightforwardly from $P(\partial G/\partial X)$.

   


\begin{thebibliography}{33}
\bibitem{comment} We chose $X_c$ to agree with other pumping
studies. However, notice that $X_c^{\rm usual} = \pi X_c$ is
customarily found in the literature instead.
\end{thebibliography}
\end{document}